\documentclass[journal=jacsat,manuscript=article]{achemso}
\setkeys{acs}{articletitle = true}

\usepackage{amsmath}
\usepackage{float}

\title{Stability and electronic properties of two-dimensional gallium}

\author{Alex Kutana}
\author{Qiyuan Ruan}
\author{Jun-Jie Zhang}
\author{Evgeni S. Penev}
\author{Boris I. Yakobson}
\email{biy@rice.edu}
\affiliation[Rice]
{Department of Materials Science and NanoEngineering, Rice University, Houston, Texas 77005, United States}

\date{}

\newcommand{\angstrom}{\mbox{\normalfont\AA}}

\begin{document}
\sloppy
\maketitle

\begin{abstract}
Two-dimensional metals offer intriguing possibilities to explore metallicity and other related properties in systems with reduced dimensionality. Here, following recent experimental reports of synthesis of two-dimensional metallic gallium (gallenene) on insulating substrates, we conduct a computational search of gallenene structures using the Particle Swarm Optimization algorithm, and identify stable low energy structures. Our calculations of the critical temperature for conventional superconductivity yield values $\sim 7$ K for gallenene. We also emulate the presence of the substrate by introducing the external confining potential and test its effect on the structures with unstable phonons.
\end{abstract}

Since their conception, two-dimensional (2D) materials have been of great fundamental and technological interest.\cite{Ferrari15,Novoselov16}
Early 2D materials were envisaged as easily exfoliable, weakly coupled 2D layers with strong in-plane binding (e.g., graphene, transition metal dichalcogenides). More recently, however, there has been growing interest in exploring the possibility of two-dimensional forms where layering is not typical in the bulk form. One of the prominent examples of such a material is monoelemental boron, where none of the natural bulk phases are layered, but which can exist in the 2D form.\cite{Mannix15,Feng16}
Other examples include silicene, germanene, and stanene.\cite{Bhimanapati15}
Recently, another monoelemental 2D material composed of gallium (gallenene), has been isolated experimentally by solid-melt exfoliation onto silica,\cite{kochat2018atomically} as well as grown epitaxially on the Si(100) and Si(111) substrates.\cite{tao2018gallenene,Li19} Located two rows down from boron in the periodic table, gallium in its 2D form is metallic, similar to borophene~\cite{Penev12,Penev16} and other monoelemental 2D metals.~\cite{Nevalaita18} Unlike borophene, however, which was originally grown on another metal (silver), gallenene was obtained on an insulating substrate, allowing exploration of the interesting properties of 2D metals with minimal interference from the substrate.

Motivated by these advances, we conduct a computational exploration of the structural stability and properties of 2D gallium. Unlike previous studies,~\cite{kochat2018atomically,Steenbergen19,Nakhaee19} which looked at a limited number of gallenene structures, we aim at carrying out a comprehensive search employing the Particle Swarm Optimization (PSO) algorithm that allows the identification of low energy structures. 
In the structures with unstable phonon modes, we tested external potential confinement to emulate the stabilizing effect of the substrate, rather than applying an in-plane strain.
We also evaluated the critical temperature for the superconducting transition from the first principle density functional theory calculations in several gallenenes to study the effects of low dimensionality and nanoconfinement.\cite{Charnaya98,Charnaya09} 

 \begin{figure}[h]
 \centering
  \includegraphics[width=0.9\textwidth]{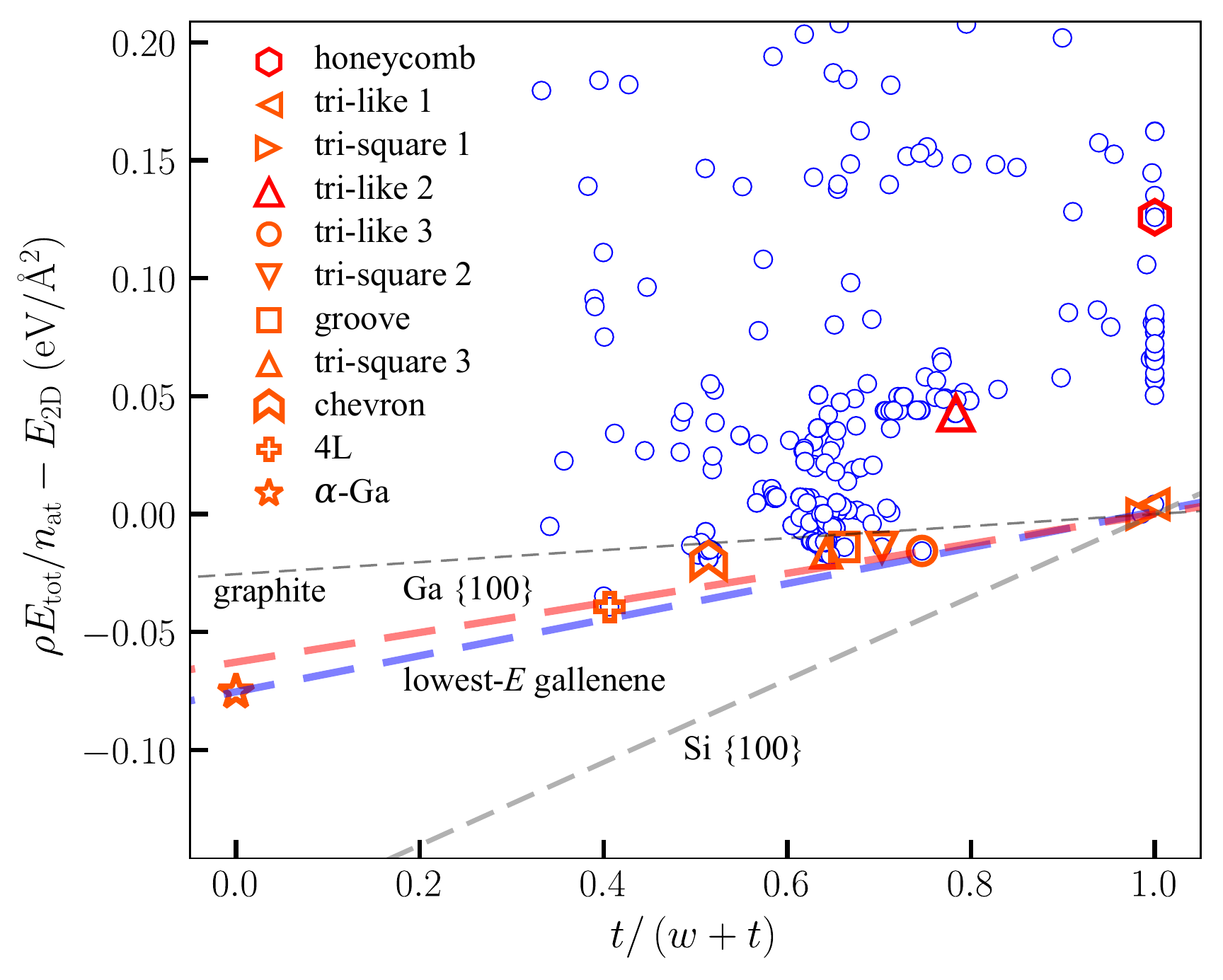}
   \caption{Total energies of gallenene structures obtained from the search. Total energy per unit area per layer relative to the most stable single-layer structure as a function of $1/n\equiv t/\left(w+t\right)$, the reciprocal of the number of layers, as defined in the text. Several low-energy structures as well structures from ref~\citenum{kochat2018atomically} are labeled. The line connecting the lowest energy single-layer gallium structure and bulk $\alpha$-Ga is drawn to show the trend, and compared with lines based on surface energies of graphite, Ga \{100\}, and Si \{100\}. Lowest-energy single-layer Ga structure is used as a reference $E_{\rm 2D}$.\label{fig:search}}
 \end{figure}

The CALYPSO code\cite{Wang12} based on the PSO algorithm has been used extensively for predicting two-dimensional materials,\cite{Wu12,LiMing15,Feng17} and is adopted here to explore gallenene.
Initially, 
random structures with certain symmetries are constructed with atomic coordinates generated by the crystallographic symmetry operations. 
To compare formation energies, structural optimizations are performed using the Quantum ESPRESSO code.\cite{Giannozzi09} 
In each generation, 70\% of the structures with lowest energy are delivered into the next generation, while the rest of the structures are randomly generated.
The number of generation is typically set to $\sim 30$, and the corresponding population size to $\sim 20$ in every generation.\cite{Wang12} 
Here, 30 generations of gallenene structures were used, each with population size 20, yielding 600 structures in total.
After the search, the structures with lowest energies are identified, as shown in Figure~\ref{fig:search}.
Among them, the structures with lowest energies were further investigated.

 \begin{figure}[h]
   \includegraphics[width=\textwidth]{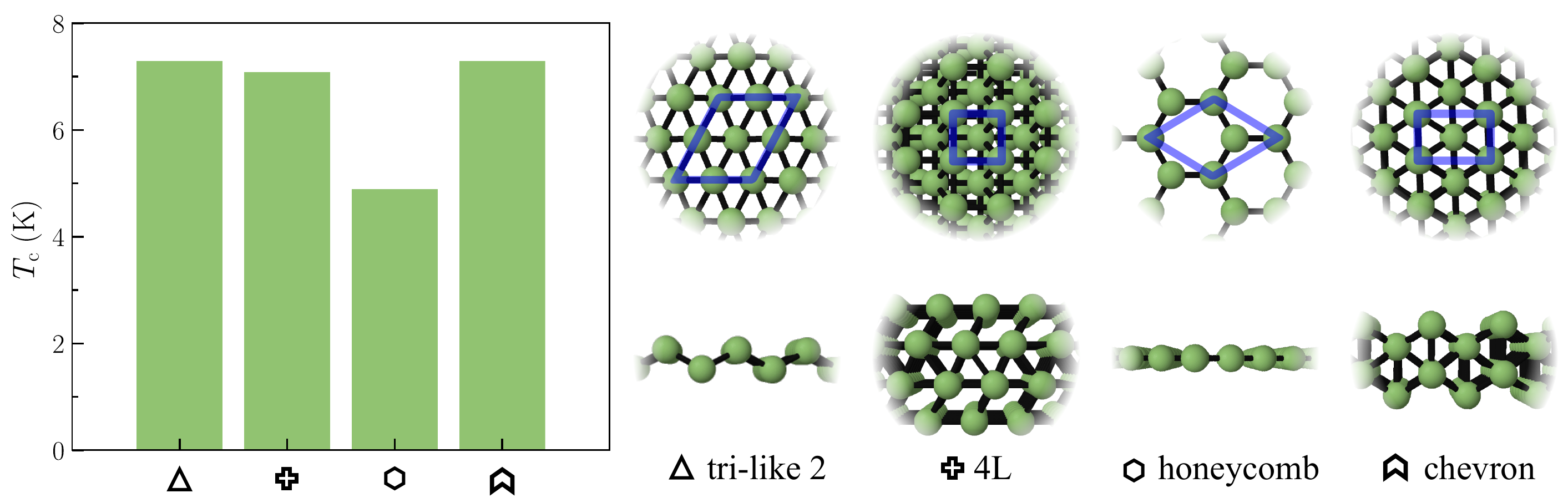}
   \caption{Calculated critical temperatures $T_{\rm c}$ for phonon-mediated superconductivity in several gallenene structures. The structures for which $T_{\rm c}$ was evaluated are shown on the right.\label{fig:Tc}}
 \end{figure}

The total energy $E_{\rm tot}$ of a layered material as a function of the number of layers $n$ can be well approximated with the following expression:
\begin{equation}
E_{\rm tot}\left(n\right)/A=nE_{\rm 2D}+\left(n-1\right)E_{\rm bind}
\label{eq:eq1}
\end{equation}
where $E_{\rm 2D}$ is the total energy of a single layer per unit area, $E_{\rm bind}$ is the binding energy between layers, and $A$ is the area of the 2D cell. 
Here, a similar expression is adopted to bulk-like materials such as Ga.
Substituting $n-1=w/t$ and $nA=n_{\rm at}/\rho$, where $w$ is the slab width (defined as the difference between the maximum and minimum of its atomic coordinates normal to layers), $t$ is the thickness per single layer, $n_{\rm at}$ is the total number of atoms in the system, and $\rho$ is the areal atom density, one obtains the following expression for the relative energy per unit area per layer:
\begin{equation}
\frac{E_{\rm tot}\left(w\right)\rho}{n_{\rm at}}-E_{\rm 2D}=E_{\rm bind}-\frac{t}{w+t}E_{\rm bind}
\label{eq:eq2}
\end{equation}
For Ga, the layer thickness $t$ was set to $4.4 \; {\text \AA}$, the lattice constant of bulk $\alpha$-Ga, giving $\rho=0.24$~\AA$^{-2}$.
To estimate the "degree of 2D-ness" of gallenene, we plot in Figure~\ref{fig:search} the left part of eq \ref{eq:eq2} for the 2D Ga structures as a function of $1/n \equiv t/\left(w+t\right)$ and compare with both 2D-like and bulk-like prototypical materials such as graphite\cite{Mounet18} and silicon.\cite{Waele16}
We also plot the line based on the theoretical surface energy\cite{Waele16} of the lowest-energy Ga surface, Ga \{100\}.
The line connecting the lowest energy single-layer gallium structure and bulk $\alpha$-Ga is also drawn, with its slope yielding the effective binding energy.
Note that the interlayer binding energy $E_{\rm bind}$, i.e. the energy of cohesion between the two layers, and the conventionally defined surface energy  $\gamma$ are related through $E_{\rm bind}\approx 2\gamma$.

One can see that in gallenene the energy trend with thickness is quite close to one that would be obtained by cleaving at the lowest-energy surface, and actually shows a higher slope, i.e. atomically thin Ga displays a bulk-like behavior.
The effective binding energy obtained from the slope is $76  \; {\rm meV /\angstrom}^2$, to be compared with $26  \; {\rm meV /\angstrom}^2$ for graphite,\cite{Mounet18} $63  \; {\rm meV /\angstrom}^2$ for Ga \{100\}, and $176  \; {\rm meV /\angstrom}^2$ for Si \{100\}.~\cite{Waele16}
The energies of the previously considered\cite{kochat2018atomically} gallenene structures, $a_{100}$ and $b_{010}$, here named "honeycomb" and "tri-like 2," respectively, are also shown in Figure~\ref{fig:search}.
Allowing out of plane displacements in the honeycomb $a_{100}$ structure leads to a triangular-like structure with 6-fold coordination, with energy similar to that of "tri-like 2."
The large slope may reflect the tendency of freestanding gallenene sheets to assume bulk form.
On the other hand, low surface energy of bulk Ga may facilitate exfoliation, although substrate and/or other kinds of confinement may play an important role in obtaining atomically thin layers.

We have calculated the electron-phonon coupling and evaluated critical temperatures $T_{\rm c}$ for phonon-mediated superconductivity in several freestanding gallenene structures, as given by the McMillan equation.\cite{McMillan68} 
The evaluation of the critical temperature is based on the microscopic theory of Bardeen, Cooper, and Schrieffer (BCS),\cite{BCS57} with the rigorous treatment of electron-phonon interactions introduced by Migdal\cite{Migdal58} and Eliashberg.\cite{Eliashberg60}
Phonon frequencies and electron-phonon coupling coefficients were calculated using the density-functional perturbation theory. 
The $T_{\rm c}$ values were obtained from the analytical approximation given by the McMillan equation,\cite{McMillan68} further modified by Allen and Dynes:\cite{Allen75} 
\begin{equation}
k_{\rm B}T_{\rm c} = \frac{\hbar \omega_{\ln}}{1.2}
\exp\left(-\frac{1.04 (1 + \lambda)}{\lambda - \mu^*-0.62\lambda\mu^*}\right)
\end{equation}

The prefactor $\omega_{\ln}$ is the logarithmically averaged phonon frequency and the effective electron-electron repulsion  was treated as an empirical parameter with the value  $\mu^*= 0.1$.\cite{Allen83,Marsiglio08}
The obtained  values of $T_{\rm c}$ are in the 4-8~K range, as shown in Figure~\ref{fig:Tc}, where the corresponding structures are also displayed.

Structural instabilities are common in freestanding atomically-thin 2D materials, manifesting as imaginary frequencies in the phonon dispersions obtained with DFT.
Using supercells as well as applying small in-plane strains are some of the useful approaches to stabilizing the phonons in such cases.
However, larger cells and extra optimization are time consuming, while strains on cell cannot imitate the stabilization effects from the substrate, as these strains are applied in the plane of the 2D materials ($xy$), but the role of the substrate is much like a confinement in the $z$ direction.
These confinement effects from the substrate can however be mimicked by applying an external potential.

\begin{figure}[h]
\includegraphics[width=0.5\columnwidth]{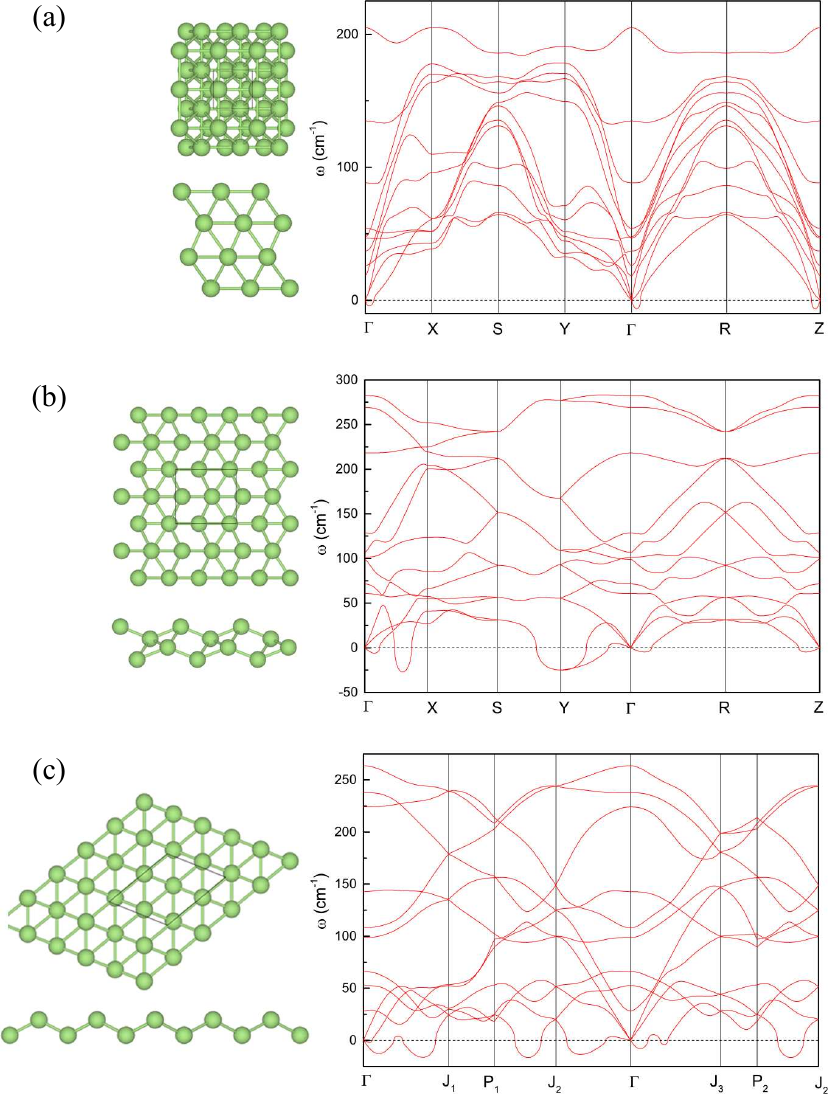}
\caption{Geometries and phonon spectra for some 2D gallenene structures obtained from the search for (a) 4-layer and (b), (c) 2-layer structures.
Figures in the left column are the geometries without the external potential.
In the right column are the corresponding phonon spectra.
Note that for the 4-layer structure, the phonons are already stable without external confinement.
For 2-layer structures, imaginary frequencies appear, indicating dynamic instabilities.\label{fig:struc}}
\end{figure}

The full interaction potential between the 2D adsorbate and substrate is an unknown function of all the atomic coordinates. 
In the harmonic approximation however, one only needs to consider the second order term of the interaction.
Therefore, an external potential of parabolic form can be used to represent this interaction in phonons calculations. 
With the potential, the nuclear equations of motion (EOMs) are modified as follows.

In the adiabatic (Born-Oppenheimer) approximation and classical limit, the lattice-dynamical properties are determined from the harmonic phonon frequencies $\omega$ which are solutions of the following secular equation:
\begin{equation}
  \det \Bigg\lvert \frac{1}{\sqrt{M_IM_J}} \frac{\partial^2 E(\mathbf{R})}{\partial \mathbf{R}_I\partial \mathbf{R}_J} - \omega^2 \Bigg\rvert = 0 \label{eqn:frequency}
\end{equation}
Here $\mathbf{R}_I$ is the coordinate of the $I$th nucleus, $M_I$ is its mass, $\mathbf{R}\equiv \{\mathbf{R}_I\}$ is the set of all the nuclear coordinates, and $E(\mathbf{R})$ is the adiabatic potential energy surface.
For monoelemental compounds such as gallenene $M_1 = M_2 = ... = M$.
After applying the external potential, the total energy $E_{\rm tot}$ becomes

\begin{equation}
  E_{\rm tot}(\mathbf{R}) = E(\mathbf{R})+E_{\rm ext}(\mathbf{R})\label{eqn:energy}
\end{equation}
Here, $E(\mathbf{R})$ is the energy of the system without the external potential and $ E_{\rm ext}(\mathbf{R})$ is the energy of the external force field. 
A quadratic confining potential of the following form is applied:

\begin{equation}
  E_{\rm ext}(\mathbf{R}) =  \frac{1}{2}k\sum_I R_{I3}^2 \label{eqn:energy1}
\end{equation}
Here, $k$ is the force constant and $R_{I3}$ is the coordinate in z-direction of the $I$th nucleus. The corresponding force on the $I$th nucleus is given by

\begin{equation}
  \mathbf{F}_{I_{\rm ext}} = -\frac{\partial E_{\rm ext}(\mathbf{R})}{\partial \mathbf{R}_I} = -k\sum_I R_{I3}\hat{\mathbf{z}}\label{eqn:force}
\end{equation}
and constant $k$ is added to the appropriate elements of the energy Hessian matrix. 
Based on the above equations, energy, force, and phonon subroutines of the Quantum ESPRESSO code were modified to implement the quadratic external potential, by adding terms in eq~\ref{eqn:energy1}, \ref{eqn:force}, and force constant $k$ to the corresponding variables in the code.

 \begin{figure}[h]
   \includegraphics[width=0.8\columnwidth]{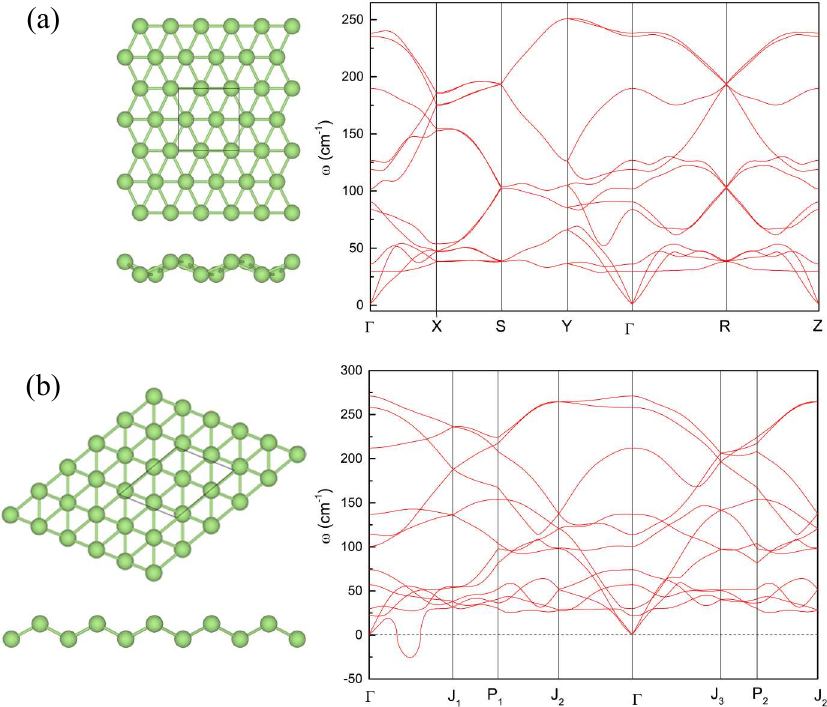}
   \caption{Geometries and phonon spectra of 2-layer gallenenes in external confinement. (a) and (b) are the structures which are also shown in Figures~\ref{fig:struc} b and c, respectively. Both structures flatten after being put into the external potential. During optimization, in-plane lattice constants, angles, and atom positions are all relaxed. Notice that the ZA branches are all shifted to $30\;\rm{cm}^{-1}$. In the second structure, imaginary frequencies for in-plane phonon modes remain.\label{fig:phon}}
 \end{figure}

In this work, a spring constant $k=3.7\;\rm{kg/s}^2$ was used.
This value corresponds to the harmonic oscillator frequency of $30\;\rm{cm}^{-1}$ for the gallium atom, which is also the frequency shift of the ZA branch at the Gamma point.
The strength of this stabilizing potential should not exceed the interaction strength between gallenene and real substrates.
Our DFT calculations show that this condition should hold for most substrates.
For example, growth of gallenene on the Si(111) substrate has been reported recently.\cite{tao2018gallenene}
We carried out a calculation of gallenene on the Si(100) substrate and obtained the second order interaction term  $k\sim 56\;\rm{kg/s}^2$, which is $\sim 15$ times larger than the value used to stabilize the gallenenes.
The interaction curve between gallenene and Si(100) substrate obtained from DFT is shown in the Supporting Information.

 The obtained phonon spectra of 2-layer structures with the external confinement are shown in Figures~\ref{fig:phon}a and b.
In the first structure, after re-optimization in the external potential, the imaginary frequencies are removed.
The second structure shows instability even in confinement, with imaginary frequencies existing between $\Gamma$ and $J_1$ points.
The atomic motions in these unstable modes are in fact in-plane, whereas the forces from the external potential act in the $z$ direction and thus can only stabilize the out-of-plane modes.
The visualization of the unstable vibrational modes can be found in Supporting Information.

In summary, we explored feasibility of two-dimensional gallium (gallenene) by performing a computational search of various two-dimensional polymorphs using the Particle Swarm Optimization algorithm. The stability trend of the structures found in the search points
at the bulklike behavior in the few-layer gallium.
At the same time, the search identified dynamically stable freestanding few-layer gallenene structures that may exist even without the stabilizing effects of the substrate.
Calculations of the critical temperature for conventional superconductivity has yielded values $\sim 7$ K in gallenene.
Finally, external potential imitating the confining effect of the substrate has been applied to the structures that displayed unstable phonon modes.
The potential is shown to be instrumental in stabilizing phonon modes with out of plane atomic displacements.

\section*{Acknowledgements}
This work was supported by the  Department of Energy BES, (Grant No. DE-SC0012547 (structural search analysis part). This research used resources of the National Energy Research Scientific Computing Center (NERSC), supported by the DOE Office of Science under Contract No. DE-AC02-05CH11231, and resources provided by the NOTS cluster at Rice University acquired with funds from NSF grant CNS-1338099.

\begin{suppinfo}
Interaction potential for gallenene on Si(111) substrate, visualizations of out-of-plane and in-plane phonon modes.
\end{suppinfo}

\bibliography{main}

\end{document}